%% file: holo-cos-14-G.tex
\documentclass[12pt]{article}
\usepackage{graphicx}
\usepackage{subfigure}
\usepackage{latexsym}

\usepackage{color}
\input{macros.tex}

\newcommand{\gs}{\mbox{$g_s$}}            
\newcommand{\ap}{\mbox{$\alpha^\prime$}}  
\newcommand{\ls}{\mbox{$l_s$}}            

\def\p{\partial}

\newcommand{\beq}{\begin{equation}}
\newcommand{\eeq}{\end{equation}}
\newcommand{\bea}{\begin{eqnarray}}
\newcommand{\eea}{\end{eqnarray}}

\begin{document}
\begin{titlepage}
\begin{flushleft}
       \hfill                     \\
       \hfill                       FIT HE - 12-04 \\
\end{flushleft}
\vspace*{3mm}
\begin{center}
{\bf\LARGE Holographic Friedmann Equation\\
 \vspace*{5mm}
  and $\cal{N}=$4 SYM theory}

\vspace*{5mm}
\vspace*{12mm}
{\large Kazuo Ghoroku\footnote[2]{\tt gouroku@dontaku.fit.ac.jp} and 
Akihiro Nakamura\footnote[4]{\tt nakamura@sci.kagoshima-u.ac.jp}
}\\
\vspace*{2mm}

\vspace*{2mm}

\vspace*{4mm}
{\large ${}^{\dagger}$Fukuoka Institute of Technology, Wajiro, 
Higashi-ku}\\
{\large Fukuoka 811-0295, Japan\\}
\vspace*{4mm}
{\large ${}^{\S}$Department of Physics, Kagoshima University, Korimoto1-21-35,Kagoshima 890-0065, Japan\\}

\vspace*{10mm}
\end{center}

\begin{abstract}

According to the AdS/CFT correspondence, 
the ${\cal N}=4$ supersymmetric Yang-Mills (SYM) theory has been studied by solving the dual
supergravity. In solving the bulk Einstein equation, we find that it 
could be related to the 4D Friedmann equation, which is solved by using the 
cosmological constant and the energy density of
the matter on the boundary, and they are dynamically decoupled from the SYM theory. 
We call this combination of the bulk Einstein equations and  
the 4D Friedmann equation as holographic Friedmann equations (HFE). 
Solving the HFE, it is shown how 
the 4D decoupled matter and the cosmological constant control the dynamical properties  of the SYM theory,
quark confinement, chiral symmetry breaking, and baryon stability.
From their effect on the SYM, {the various kinds of matter} are separated to two groups.
Our results would give important information in studying the cosmological development of our universe. 
 
\end{abstract}
\end{titlepage}

\section{Introduction}

Up to now, various holographic approaches to the ${\cal N}=4$ supersymmetric 
Yang-Mills (SYM) theory with strong coupling
have been performed from the dual supergravity \cite{MGW}-\cite{CNP}. 
In these, the research has been extended 
to the SYM in the background $dS_4 (AdS_4)$ by introducing 4D cosmological constant
($\Lambda_4$) \cite{Hawking,Alishahiha1,Alishahiha2,H,GIN1,GIN2,EGR,EGR2}. 
Then, it has been found that 
the dynamics of the SYM theory is largely controlled by
the 4D geometry, the dS$_4$ and AdS$_4$ \cite{GIN1,GIN2}. 

In the case of dS$_4$ ($\Lambda_4>0$), a horizon 
\footnote{Here we express the horizon by the zero points of the metric, which is not necessarily
the time component.} 
exists in the bulk geometry as in the case of 
AdS$_5$-Schwarzschild background, which is dual to the SYM theory in
the high temperature deconfinement phase. As expected from this similarity of the background, 
the gauge theory in dS$_4$ is in the deconfinement phase even if the theory is in the confinement
phase at the limit of $\Lambda_4=0$  \cite{GIN1}. In fact, the positive
$\Lambda_4$ plays a role similar to the temperature, then we could see the screening of 
the confining force above the corresponding scale of $\Lambda_4$.

{For $\Lambda_4=0$, the 4D boundary background is represented by the Minkowski
space-time, and the confinement phase is realized by introducing a non-trivial dilaton, which
implies the condensate of the gauge field strength.
This condensate provides the tension of the 
linear potential between the quark and the anti-quark  \cite{GY}.
In the present universe, however, 
very small positive $\Lambda_4$ would be believed to exist, then the cofining force responsible to dilaton
would be screened.
However the screening effect would appears at very large distance between quarks, so we 
could fortunately find stable mesons
and baryons in our universe since the effect of $\Lambda_4$ is negligible within the hadron scale.
However, at early universe, the situation would be changed.}

On the other hand,
in the case of AdS$_4$ ($\Lambda_4<0$), the horizon disappears and we could find both
the quark confinement and chiral symmetry breaking even if we neglect the effect of the non-trivial dilaton, 
which is necessary in the Minkowski space-time for the confinement. \footnote{It would be interesting to compare this result with the observation given in
\cite{AIS} many years ago. The authors in \cite{AIS} has found a discrete mass
spectrum for a free scalar field in AdS$_4$, and this spectrum coincides with our result for
the mass spectrum of a scalar meson, which is considered as a bound state of quark and
anti-quark in our holographic approach.}
In  dS$_4$, however, the 
screening effect of the positive $\Lambda_4$ overwhelms the dilaton effect at large scale. 
{In the AdS$_4$,} we find that
the negative $\Lambda_4$ and the dilaton are cooperative to realize the confinement,
and furthermore the negative $\Lambda_4$ induces the chiral symmetry breaking  \cite{GIN2}.

\vspace{.3cm}
In these approaches, we found that the deformation of the boundary space-time due to the
$\Lambda_4$ plays an important role in determining the dynamical properties of the SYM theory living
in this curved space-time. 
{This point is understood from the bulk metric which is also deformed
from the AdS$_5$, and here we denote it as $\widetilde{AdS_5}$. An explicit
example of such a $\widetilde{AdS_5}$ is shown in the Sec. 4 by the Eqs. (\ref{10dmetric-23})-(\ref{sol-11-3}), 
\footnote{The Eqs. (\ref{10dmetric-2})-(\ref{sol-12}) in the Sec. 2 also represent another kind of
$\widetilde{AdS_5}$}
and we should notice that it is reduced to the undeformed AdS$_5$ in the limit of $\lambda=0$.
The situation
is the same with the case of AdS$_5$-Schwarzschild background which reduces to AdS$_5$ at the
zero temperature limit. Thus the deformation of $\widetilde{AdS_5}$ is charcterized
by the parameter(s) of the boundary theory, e.g. temperature, 4D cosmological constant, e.t.c..
These parameters are essential to determine the dynamical properties of the CFT on the boundary.
Thus, as a result of the analyses up to now, it would be able to extend the duality relation as follows,
\beq\label{duality-ex}
         {\rm SUGRA~in}~ \widetilde{AdS_5}~ \leftarrow {\rm dual} \rightarrow ~{\rm CFT ~in~4D~curved~space~time},
\eeq
where the bulk
$\widetilde{AdS_5}$ comes back to the AdS$_5$ by reducing the parameter(s) of the 4D theory living
on the boundary, and we find
\beq\label{duality-ori}
         {\rm SUGRA~in}~ {\rm{AdS}_5}~ \leftarrow {\rm dual} \rightarrow ~{\rm CFT ~in~4D~Minkowski~space~time},
\eeq

}

\vspace{.3cm}
According to the above idea, we extends our analysis to the case of time dependent $\lambda$
by introducing various kinds of matter living in the boundary.
In our universe, there are many other ingredients, which control the 4D space-time of
our universe, other than $\Lambda_4$. 
It is therefore important to make clear how do they could change the properties of the SYM
theory. This issue is examined here by extending our holographic analysis to the case where 
various kinds of matter are included. This is performed
simply by replacing the $\Lambda_4$ to a time dependent form, $\lambda(t)$, which is given as
\beq\label{lambda}
   \lambda(t)={1\over 3}\left(\Lambda_4+\kappa^2\sum_u {\rho_u\over {a_0{(t)}}^{3(1+u)}}\right)\equiv
   \sum_{n=3u+2}\lambda_n\, ,
\eeq
where $\kappa^2$ and $a_0(t)$ denote the 4D gravitational constant and the three dimensional scale factor of
the Robertson-Walker metric, 
\beq\label{RW-0}
  ds_{(4)}^2=-dt^2+a_0(t)^2\gamma_{ij}dx^idx^j\, ,
\eeq
which is set as the boundary metric in our analysis. {Here $\gamma_{ij}=\delta_{ij}/(1+k\sum_i {x^i}^2)^2/4$ and
$k=\pm 1,$ or $0$.}
The energy density of the various kinds of
matter are expressed by $\rho_u$, where $u$ denotes the ratio of the pressure $p$ to the energy density
of the matter,
\beq
  u={p\over \rho_u}\, .
\eeq
The $\lambda_n$ of the right hand side is introduced to describe each term in the form of
$\lambda_n\propto 1/a_0(t)^{n+1}$, where $n=3u+2$. 
{In this notation, we could express various kinds of
ordinary matter by integer $n$, however we extend to the case
of non-integer in order to include abnormal kinds of matter.}

\vspace{.3cm}
The generalized $\lambda(t)$ appears in the bulk Einstein equations, 
then the differential equation for the dilaton becomes a little complicated \cite{EGR2}
due to the time-dependence of the $\lambda$. 
{So we set the dilaton as a trivial
one for the simplicity since our purpose is to make clear 
the role of the various kinds of matter as in the case that we have studied the effects of $\Lambda_4$ on the SYM theory.
As for the competition with the dilaton contribution, it will be postponed to study it into the future.}

When  $\lambda$ depends on the time, the form of the time component metric ($g_{00}$) is 
largely changed from the
one of the time-independent case of $\Lambda_4$. On the other hand, other components are not changed in the form. Due to
this modification of the metric, it is not so simple to see the changing of the dynamical properties of SYM 
from the case of the constant $\Lambda_4$. By picking up one component $\lambda_n$ from $\lambda(t)$, its effect
on the SYM theory is studied, and we find that many kinds of matter are classified to two groups, (A) $\lambda_n<0$ and $n<0$, and
(B) others. For group (A), quark confinement and chiral symmetry breaking are realized, and
negative $\Lambda_4$ is included in this group. On the other hand, positive $\Lambda_4$ and the
ordinary kinds of matter are included in the group (B), and the quark deconfinement phase is realized. As for the chiral
symmetry, it is restored for most cases except for very large $n$ case. We discuss about the possibility of
this exotic matter of large $n$ and positive $\lambda_n$.

Furthermore, we examined the effect of the matter on the baryon vertex which is expressed by the
D5 brane in the present type IIB model. The D5 brane wraps on $S^5$, then its action depends only on 
$g_{00}$ of (A)d$S_5$ and the metric of $S^5$. As a result, we can see the effect of $g_{00}$ 
directly in this case,  and we find the stability of the vertex in both groups when the condition,
$n\lambda_n>0$, is satisfied.

\vspace{.5cm}
In the next section, the model for the bulk theory is given and the holographic Friedmann equation
is explained. Then the background solution used here is obtained. In the section 3, we examine 
the energy momentum tensor of holographic SYM theory in order to see the relation to the one
of the matter. 
In the section 4, the role of various kinds of matter in determining the dynamical properties of the
SYM is examined through the Wilson loop and the chiral condensate by introducing the probe D7 brane. 
In the section 5, the stability of the baryon vertex is examined. 
Summary and discussions are given in the final section.

\section{Setup of holographic theory}

Firstly we briefly review our model.
We start from the 
10d type IIB supergravity retaining the dilaton
$\Phi$, axion $\chi$ and selfdual five form field strength $F_{(5)}$,
\beq\label{2Baction}
 S={1\over 2\kappa^2}\int d^{10}x\sqrt{-g}\left(R-
{1\over 2}(\partial \Phi)^2+{1\over 2}e^{2\Phi}(\partial \chi)^2
-{1\over 4\cdot 5!}F_{(5)}^2
\right), \label{10d-action}
\eeq
where other fields are neglected since {we do not need} them, and 
$\chi$ is Wick rotated \cite{GGP}.
Under the Freund-Rubin
ansatz for $F_{(5)}$, 
$F_{\mu_1\cdots\mu_5}=-\sqrt{\Lambda}/2~\epsilon_{\mu_1\cdots\mu_5}$ 
\cite{KS2,LT}, and for the 10d metric as $M_5\times S^5$ or
$$ds^2_{10}=g_{MN}dx^Mdx^N+g_{ij}dx^idx^j=g_{MN}dx^Mdx^N+R^2d\Omega_{5}^2\, ,$$ 
we consider the solution. Here, the parameter is set as $(\mu=)1/R=\sqrt{\Lambda}/2$.

The equations of motion of non-compact five dimensional part
$M_5$ are written as
\footnote{The five dimensional $M_5$ part of the
solution is obtained by solving the following reduced 
Einstein frame 5d action,
\beq\label{action}
 S={1\over 2\kappa_5^2}\int d^5x\sqrt{-g}\left(R+3\Lambda-
{1\over 2}(\partial \Phi)^2+{1\over 2}e^{2\Phi}(\partial \chi)^2
\right), \label{5d-action}
\eeq
which is written 
in the string frame and taking $\alpha'=g_s=1$ and the opposite sign
of the kinetic term of $\chi$ is due to the fact that
the Euclidean version is considered here \cite{GGP}.}
\beq\label{metric}
 R_{MN}={1\over 2}\left(\partial_M\Phi\partial_N\Phi-
         e^{2\Phi}\partial_M\chi\partial_N\chi\right)-\Lambda g_{MN}
\eeq
\beq \label{p}
 {1\over \sqrt{-g}}\partial_M\left(\sqrt{-g}g^{MN}\partial_N\Phi\right)=
    -e^{2\Phi}g^{MN}\partial_M\chi\partial_N\chi\ , 
\eeq 
\beq\label{chieq}
   \partial_M\left(\sqrt{-g}e^{2\Phi}g^{MN}\partial_N\chi\right)=0
\eeq
These equations have a supersymmetric solutions when the following ansatz is
imposed for the axion $\chi$ \cite{KS2,LT,GGP},
\beq\label{ansa1}
     \chi=-e^{-\Phi}+\chi_0 \ .
\eeq
And this ansatz is also useful in getting non-supersymmetric solutions. The merit
to use this ansatz (\ref{ansa1}) is to be able to reduce the above equations (\ref{metric})-
(\ref{chieq}) to the following two forms,
\beq\label{gravity}
 R_{MN}=-\Lambda g_{MN}
\eeq
and
\beq\label{phi}
   \partial_M\left(\sqrt{-g}g^{MN}\partial_N e^{\Phi}\right)=0
\eeq
where we notice that the two equations (\ref{p}) and (\ref{chieq}) are rewritten
to the same form with (\ref{phi}). So $\Phi$ and $\chi$ are obtained by using the solution of (\ref{gravity}),
however the dilaton is not important here as mentioned in the introduction. Then we set as $\Phi=\chi=0$ hereafter
for the simplicity.

\subsection{Holographic Friedmann equation}\label{sec22}

We solve the 5D Einstein equation (\ref{gravity}) 
by supposing the following metric and coordinates,
\beq
ds^2_{\rm E}=-n^2(t,y)dt^2+a(t,y)^2\gamma_{i,j}dx^idx^j+dy^2 \ . 
\label{y-co}
\eeq
for the Einstein frame metric \cite{BDEL} since
this metric is useful to study the cosmological development of the universe.

\vspace{.3cm}
\noindent {\bf Dark Radiation ($C$)}

\vspace{.3cm}
In terms of this metric, the following equation is obtained from the Einstein equation of
$tt$ and $yy$ components of
(\ref{gravity}) \cite{BDEL,Lang},
\beq\label{cosmo}
 \left({\dot{a}\over na}\right)^2+{k\over a^2}=
   -{\Lambda\over 4}+\left({a'\over a}\right)^2
  +{C\over a^4}\ ,
\eeq
where $\dot{a}=\partial a/\partial t$ and $a'=\partial a/\partial y$. {The parameter $k$ is set as $\pm 1$ or 0
according to the situation and the sign of 4D cosmological constant.}
The integration constant $C$ must be a constant with respect to both $y$ and
$t$ in order to satisfy other components of Einstein equations. 
The term proportional to $C$ is called as ''dark radiation" since it is proportional to $a^{-4}$. 
From holographic viewpoint,
it has been cleared that this term corresponds to the energy density of
the thermal Yang Mills fields with a definite temperature \cite{NS,EGR}.

\vspace{.5cm}
\noindent {\bf Holographic Friedmann Equation}

\vspace{.3cm}
Further, by setting the following ansatz \cite{BDEL,Lang}, 
\beq\label{eqn}
   n(t,y)={\dot{a}(t,y)\over \dot{a}_0(t)}\, , 
        \quad a=a_0(t) A(t,y)\, , 
\eeq
the Eqs.(\ref{cosmo}) and (\ref{eqn}) are rewritten as,
\beq\label{A}
 \left({\dot{a}_0\over a_0}\right)^2+{k\over a_0^2}=
   -{\Lambda\over 4}A^2+\left({A'}\right)^2
  +{C\over a_0^4 A^2}\ , 
\eeq
\beq
ds^2_{\rm E}=A^2(y,t)\left(-\bar{n}^2(t,y)dt^2+a_0(t)^2\gamma_{i,j}dx^idx^j\right)+dy^2 \ , 
          \quad
   \bar{n}={n\over A}\, . \label{5d-metric-2}\, 
\eeq
At this stage, there are three unknown functions, $a_0(t)$, $A(y,t)$ and $n(y,t)$, 
in spite of the two equations,
(\ref{eqn}) and (\ref{A}), to be solved.
Then one of the three should be given by some reasonable assumption.
Our strategy is to find $A(y,t)$ and $n(y,t)$ by solving (\ref{eqn}) and (\ref{A}) 
with $a_0(t)$ which is given as a solution of the equation on the boundary. 

\vspace{.3cm}
Here the boundary of the present bulk is represented by the following metric
\beq\label{RW}
  ds_{(4)}^2=-dt^2+a_0(t)^2\gamma_{ij}dx^idx^j\, ,
\eeq
since our solution behaves as $\bar{n}\to 1$ and $A(y,t)\to \infty$ for $y\to\infty$ as shown 
{in the next section.} Then 
the 4D Friedmann equation appeared in the standard cosmology is expressed as
\bea\label{bc-RW2}
  \left({\dot{a}_0\over a_0}\right)^2+{k\over a_0^2}&=& 
 {\Lambda_4\over 3}+{\kappa_4^2\over 3}\left(
    {\rho_m\over a_0^3}+ {\rho_r\over a_0^4}+{\rho_u\over a_0^{3(1+u)}} \right)=
        \lambda(t)\, \label{bc-RW3}
\eea
where $\kappa_4$ ($\Lambda_4$) denotes the 4D gravitational constant (cosmological constant).  $\rho_m$ and $\rho_r$ denote the energy density of 
the nonrelativistic matter and the radiation of 4D theory respectively. The $\lambda(t)$ in (\ref{bc-RW3})
represents the same one of (\ref{lambda}) given above although some terms are written explicitly in the 
equation (\ref{bc-RW3}). It is important to be able to solve the bulk equation (\ref{A}) by relating
its left hand side to the Friedmann equation on the boundary at any $y$.

\vspace{.3cm} 
In this way, $A(y,t)$ and $n(y,t)$ are solved by using the time dependent function
$\lambda(t)$. We notice here that, in solving for $A(y,t)$ and $n(y,t)$, it is not necessary to know the 
explicit form of $\lambda(t)$. As for the explicit form of $a_0(t)$, we discuss {it} in the final
stage by solving (\ref{bc-RW2}) for $a_0(t)$. Through this procedure, we can see how the 
parameters of 4D Friedmann equation  (\ref{bc-RW2}) control the 4D dual theory, the SYM theory.
Further we notice following points;
In a sense, the Eq.(\ref{A}) is similar to (\ref{bc-RW2}) given at the boundary. So we could
interpret the Eq.(\ref{A}) as the
Friedmann equation given at the slice of finite $y$ off the boundary. In this sense, 
we call Eq.(\ref{A}) here as the ''{holographic Friedman equation}" by combining it 
with the 4D Friedmann equation (\ref{bc-RW2}).

\vspace{.5cm}
Here we must be careful about the relation between the radiation density $\rho_r$ and the dark radiation $C$.
By comparing Eqs.(\ref{bc-RW2}) and (\ref{A}), 
one might expect that the dark radiation term may
approach to the radiation term of  (\ref{bc-RW2}) in the boundary limit as
\beq
   {C\over A^2}\mathop{\to }_{y\to\infty}  {\kappa_4^2\over 3}{\rho_r}\ \label{4DG} .
\eeq
However, this correspondence is misleading since $1/A^2\to 0$ at the boundary. Then $ {C\over A^2}$ disappears
there. This fact is consistent
with the fact that the dark radiation belongs to the SYM theory, which is dual to the bulk gravity. 
So it should decouple from the gravity on the boundary, and then it doesn't appear in the 
4D Freedmann equation of the boundary. This is assured from the fact that
the effective gravitational coupling constant is expressed by $1/A^2$ which vanishes at the boundary.
So ${\rho_r}$ in Eq.(\ref{bc-RW2}) has
nothing to do with the dark radiation. The radiation $\rho_r$ in 4D boundary is 
therefore independent of the dual SYM theory.
However, we must notice that both radiations give 
dynamical effects on the holographic gauge theory.

\vspace{.3cm}
\noindent{\bf Solution}

Finally, in this section, we give
the solution of $A$ and $n$ by using $\lambda(t)$. They are obtained by 
replacing the coordinate from $y$ to $r$ defined as
$r/R=e^{\mu y}$. Then, from (\ref{eqn})-(\ref{5d-metric-2}), we have 
\beq\label{10dmetric-2}
ds^2_{10}={r^2 \over R^2}\left(-\bar{n}^2dt^2+\bar{A}^2a_0^2(t)\gamma^2(x)(dx^i)^2\right)+
\frac{R^2}{r^2} dr^2 +R^2d\Omega_5^2 \ . 
\eeq
\bea
 \bar{A}&=&\left(\left(1-{\lambda\over 4\mu^2}\left({R\over r}\right)^2\right)^2+\tilde{c}_0 \left({R\over r}\right)^{4}\right)^{1/2}\, , \label{sol-10} \\
\bar{n}&=&{\left(1-{\lambda\over 4\mu^2}\left({R\over r}\right)^2\right)
         \left(1-{\lambda+{a_0\over \dot{a}_0}\dot{\lambda} \over 4\mu^2}\left({R\over r}\right)^2\right)-\tilde{c}_0 \left({R\over r}\right)^{4}\over 
       \sqrt{\left(1-{\lambda\over 4\mu^2}\left({R\over r}\right)^2\right)^2+\tilde{c}_0 \left({R\over r}\right)^{4}}}\, , \label{sol-11}
\eea
where 
\beq
\tilde{c}_0=C/(4\mu^2a_0^4)\, . \label{sol-12}
\eeq
We can see that the boundary geometry coincides with (\ref{bc-RW2}) since $\bar{A}$ and $\bar{n}$ approach to one
for $r\to\infty~ (y\to \infty)$. Then (\ref{bc-RW2}) can be used as a boundary condition, and it determines $a_0(t)$.


\vspace{.3cm}
\section{Energy Momentum $\langle T_{\mu\nu}\rangle$}

At first, we study 4D stress tensor from holographic approach. %
In order to give it, we rewrite
the 5d part of the metric (\ref{10dmetric-2}).
According to the Fefferman-Graham framework \cite{KSS,BFS,FG}, it is given as
\bea\label{thermal}
 ds^2_{(5)}&=&{r^2 \over R^2}\left(-\bar{n}^2dt^2+\bar{A}^2a_0^2(t)\gamma^2(x)(dx^i)^2\right)+
\frac{R^2}{r^2} dr^2 \nonumber \\
&=& {1\over \rho}\hat{g}_{\mu\nu}dx^{\mu}dx^{\nu}+
{d\rho^2\over 4\rho^2}={1\over \rho}\left(-\bar{n}^2dt^2+\bar{A}^2a_0^2(t)\gamma^2(x)(dx^i)^2\right)+
{d\rho^2\over 4\rho^2}
\eea
where $\rho=1/r^2$, $R=1$ and
\bea
 \bar{A}&=&\left(\left(1-{\lambda\over 4\mu^2}\left({\rho\over R^2}\right)\right)^2+\tilde{c}_0 \left({\rho\over R^2}\right)^{2}\right)^{1/2}\, , \\
\bar{n}&=&{\left(1-{\lambda\over 4\mu^2}\left({\rho\over R^2}\right)\right)
         \left(1-{\lambda+\dot{\lambda}a_0/\dot{a}_0\over 4\mu^2}\left({\rho\over R^2}\right)\right)-\tilde{c}_0 \left({\rho\over R^2}\right)^{2}\over  \bar{A}}
\eea
In the present case, $\hat{g}_{\mu\nu}$ is expanded as \cite{BFS}
\beq
  {g}_{\mu\nu}={g}_{(0)\mu\nu}+{g}_{(2)\mu\nu}\rho+
 {\rho^2}\left({g}_{(4)\mu\nu}+{h}_{1(4)\mu\nu}\log\rho
+{h}_{2(4)\mu\nu}(\log\rho)^2\right)+\cdots\, ,
\eeq\label{Feff1}
where 
\beq
 {g}_{(0)\mu\nu}=({g}_{(0)00},~{g}_{(0)ij})=(-1,~a_0(t)^2\gamma_{i,j})\, , 
\eeq
and 
\beq 
{g}_{(2)\mu\nu}={\lambda\over 2}\left(1+{{a_0\over \dot{a}_0}\dot{\lambda}\over \lambda},-~{g}_{(0)ij}) \right)\, , 
\eeq
\beq
 {g}_{(4)\mu\nu}={\tilde{c}_0\over R^4}~(3,~{g}_{(0)ij})+
                {\lambda^2\over 16}~\left(-{(\lambda+{a_0\over \dot{a}_0}\dot{\lambda})^2\over \lambda^2},~{g}_{(0)ij}\right)\, .
\eeq\label{Feff2}

\vspace{.5cm}
Then by using the following formula \cite{KSS},
\beq
 \langle T_{\mu\nu}\rangle={4R^3\over 16\pi G_N}\left({g}_{(4)\mu\nu}-
 {1\over 8}{g}_{(0)\mu\nu}\left( ({\rm Tr}g_{(2)})^2-{\rm Tr}g_{(2)}^2\right)
   -{1\over 2}\left({g}_{(2)}^2\right)_{\mu\nu}+{1\over 4}{g}_{(2)\mu\nu}
    {\rm Tr}g_{(2)}\right)\, ,
\eeq\label{Feff5}
we find 
\bea
 \langle T_{\mu\nu}\rangle &=&{4R^3\over 16\pi G_N^{(5)}}~\left({\tilde{c}_0\over R^4}~(3,~{g}_{(0)ij})+
    {3\lambda^2\over 16}~\left(1, \beta~{g}_{(0)ij}\right)  \right)\, , \\ 
\beta &=&-\left(1+{2{a_0\over \dot{a}_0}\dot{\lambda}\over 3\lambda}\right)\,\, .
\eea\label{stress0}

Then we have
\beq
 \langle T_{\mu\nu}\rangle=\langle \tilde{T}_{\mu\nu}^{(0)}\rangle+
{4R^3\over 16\pi G_N^{(5)}}\left\{{3\lambda^2\over 16}\left(1,~\beta{g}_{(0)ij}\right)\right\}\, .
\label{Feff6}
\eeq
\beq\label{Feff6-2}
   \langle \tilde{T}_{\mu\nu}^{(0)}\rangle={4R^3\over 16\pi G_N^{(5)}}
{\tilde{c}_0\over R^4}(3,~{g}_{(0)ij})\, ,
\eeq
where $\langle \tilde{T}_{\mu\nu}^{(0)}\rangle$ is the stress tensor corresponding to 
the thermal YM fields given in (\ref{stress0}) for the case of
$\lambda=0$. The second term comes from the loop corrections of the YM fields in the curved
space-time.
While the first term does not contribute to
the conformal anomaly,  namely
\beq
  \langle \tilde{T}_{\mu}^{{(0)}\mu}\rangle=0\, ,
\eeq
the second term leads to the anomaly as follows
\beq
 \langle T_{\mu}^{\mu}\rangle=-{3\lambda^2\left(1+{\dot{\lambda}\over 2\lambda}{a_0\over \dot{a}_0}\right)\over 8\pi^2}N^2\, ,\label{anomaly2}
\eeq
where we used $G_N^{(5)}=8\pi^3{\alpha'}^4g_s/R^5$ and $R^4=4\pi N{\alpha'}^2g_s$.

\vspace{.3cm}
We can see that the anomaly (\ref{anomaly2}) is the same one obtained 
from the loop corrections in the $\cal{N}=$ $4$ SYM theory
for a curved space-time, which is given by (\ref{RW})
as the boundary space-time here. In this background, the curvature squared terms, which
are responsible to the anomaly, are given as
\bea
  R^{\mu\nu\lambda\sigma}R_{\mu\nu\lambda\sigma}&=&12\left(
      2\lambda^2+{\dot{\lambda}\lambda}{a_0\over \dot{a}_0}+\left(\dot{\lambda}{a_0\over 2\dot{a}_0}\right)^2\right)\, , \\
   R^{\mu\nu}R_{\mu\nu}&=&12\left(
      3\lambda^2+3{\dot{\lambda}\lambda}{a_0\over 2\dot{a}_0}+\left(\dot{\lambda}{a_0\over 2\dot{a}_0}\right)^2\right)\, , \\
   {1\over 3}R^{2}&=&12\left(
      4\lambda^2+4{\dot{\lambda}\lambda}{a_0\over 2\dot{a}_0}+\left(\dot{\lambda}{a_0\over 2\dot{a}_0}\right)^2\right)\, .
\eea
In general, the conformal anomaly for $n_s$ scalars,
$n_f$ Dirac fermions and $n_v$ vector fields is given as \cite{birrell,Duff93}
\beq
 \langle T_{\mu}^{\mu}\rangle=-{n_s+11n_f+62n_v\over 90\pi^2}E_{(4)}
        -{n_s+6n_f+12n_v\over 30\pi^2}I_{(4)}\, ,
\eeq
\beq
  E_{(4)}={1\over 64}\left(R^{\mu\nu\lambda\sigma}R_{\mu\nu\lambda\sigma}
       -4R^{\mu\nu}R_{\mu\nu}+R^2\right)\, ,
\eeq
\beq
  I_{(4)}=-{1\over 64}\left(R^{\mu\nu\lambda\sigma}R_{\mu\nu\lambda\sigma}
       -2R^{\mu\nu}R_{\mu\nu}+{1\over 3}R^2\right)\, ,
\eeq
where $\Box R$ has been abbreviated since it does not contribute here.
For the $\cal{N}=$ $4$ SYM theory, the numbers of the fields are given by
$N^2-1$ times the number of each fields, which are 
equivalent to $n_s=6$, $n_f=2$ and $n_v=1$. Then we find for large $N$,
\beq
 \langle T_{\mu}^{\mu}\rangle={N^2\over 32\pi^2}\left(
       R^{\mu\nu}R_{\mu\nu}-{1\over 3}R^2\right)=-{3\lambda^2\left(1+{\dot{\lambda}\over 2\lambda}{a_0\over \dot{a}_0}\right)\over 8\pi^2}N^2\, .
       \label{anomaly3}
\eeq
This result (\ref{anomaly3}) is precisely equivalent to the above holographic one (\ref{anomaly2}).

{Thus we could show that the holographic analysis could provide correct results for the energy
momentum tensor even if the metric is time dependent. Then we can say that
any matter field, which decouples to the $\cal{N}=$ $4$ 
SYM theory, could give an influence to the $\cal{N}=$ $4$ SYM theory through the 
curvatures. 
}

\vspace{.3cm}
\noindent{\bf Continuity equation of  SYM fields}

Before seeing the other effects, we see another important fact. The energy density of the SYM
theory is composed of two kind of contents, the one of the thermal SYM fields and the one of 
the vacuum energy ($\lambda^2$-dependent terms) obtained as the quantum corrections. 
This density obeys the following continuity equation,
\beq
 \dot{\rho}+3H(\rho +p)=0\, ,
 \label{continuity}
\eeq
where $\rho$, $p$ and $H$ represent the energy density, pressure and the Hubble 
constant. Here $H=\dot{a}_0/a_0$.
In the present case, we obtain them from (\ref{Feff6}) as
\beq
 \rho=3 \alpha \left({\tilde{c}_0\over R^4}+{\lambda^2\over 16}\right)\, ,
 \quad p=\alpha \left({\tilde{c}_0\over R^4}-3{\lambda^2\over 16}\left(
      1+{2\dot{\lambda}\over 3\lambda}{a_0\over \dot{a}_0}\right)\right)\, ,
 \quad \alpha={4R^3\over 16\pi G_N^{(5)}}\, , \label{density}
\eeq
It is easy to see that the continuity equation (\ref{continuity}) is satisfied by the above
$\rho$ and $p$. This is satisfied even if $\lambda=0$, therefore the $\lambda$ dependent part also satisfies
without the thermal part. This fact implies that the energy momentum is not transferred between the thermal part
(SYM part) and the $\lambda(t)$ dependent part (matter part). 
{This is consistent with the fact that the SYM theory studied here decouples to the gravity and also to
the matter in the $\lambda(t)$.
In other words, the various kinds of
matter are responsible for determining $a(t)$ through the 4D Friedmann equation, 
however the SYM fields aren't.
So the role of the matter is to reform the space-time background
where the SYM theory lives.
As a result, this reformed background changes 
the dynamical properties of the SYM theory as shown below.}

\vspace{.3cm}
\section{Dynamical role of decoupled matter}

Here we study the dynamical effect of the matter on the SYM theory. Many kinds of matter are
introduced through the time-dependent cosmological
term $\lambda(t)$ as in (\ref{bc-RW2}). 
In order to see its pure effect, we set $\tilde{c}_0=0$. Then the metric is written as 
\beq\label{10dmetric-23}
ds^2_{10}={r^2 \over R^2}\left(-\bar{n}^2dt^2+\bar{A}^2a_0^2(t)\gamma^2(x)(dx^i)^2\right)+
\frac{R^2}{r^2} dr^2 +R^2d\Omega_5^2 \ . 
\eeq
\bea
 \bar{A}&=&\left(1-{\lambda\over 4\mu^2}\left({R\over r}\right)^2\right)\, , \label{sol-10-3} \\
\bar{n}&=&\left(1-{\lambda+{a_0\over \dot{a}_0}\dot{\lambda} \over 4\mu^2}\left({R\over r}\right)^2\right)\,
 , \label{sol-11-3}
\eea

The $\bar{n}$ is simplified when we restrict to one kind of matter field ($\lambda_m$) in $\lambda$ whose energy
density behaves as $\epsilon\propto 1/a_0(t)^{n+1}$. In this case, $\bar{n}$ is written as follows
\beq
  \bar{n}=\left(1+n{\lambda_m \over 4\mu^2}\left({R\over r}\right)^2\right)\,
 . \label{sol-11-4}
\eeq
We notice here the following point that, in $\bar{n}$ for $n>0$ and positive $\lambda_m$, 
the horizon in  $\bar{n}$ disappears. On the other hand, it remains in $\bar{A}$.

\vspace{.3cm}
\subsection{Wilson-Loop and Quark Confinement}

The quarks are introduced through probe D7 branes.
By presuming D7 brane embedding, we can consider the Wilson-Loop 
whose boundary is on the D7 brane. 
In order to study the potential between
quark and anti-quark, we consider the U-shaped ( in $r-x$ space) string whose two end-points are on D7 brane
as studied in \cite{GIN1}.
Supposing
the string action whose world volume is set in $(t,x)$ plane \footnote{Here $x$ denotes one of the three coordinate $x^i$, and we take $x^1$ in the present case.}, the energy $E$ 
of this state is obtained as a function of the distance ($L$) between
the qaurk and anti-quark according to \cite{GIN1}.

Taking the gauge as $X^0=t=\tau$ and $X^1=x^1=\sigma$
for the coordinates $(\tau,~\sigma)$ of string world-volume,
the Nambu-Goto Lagrangian in the present background (\ref{10dmetric-2}) 
becomes
\beq
   L_{\textrm{\scriptsize NG}}=-{1 \over 2 \pi \alpha'}\int d\sigma ~
   \bar{n}(r)\sqrt{r'{}^2
        +\left({r\over R}\right)^4 \left(\bar{A}(r)a_0(t)\gamma(x)\right)^2 
} ,
 \label{ng}
\eeq
where we notice $r'=\partial r/\partial x=\partial r/\partial \sigma$.
Then the energy of this configuration is rewritten to more convenient form
according to Gubser \cite{Gub} as
\beq
 E=-L_{\rm NG}
  ={1\over 2\pi \alpha'} \int d \tilde{\sigma} ~ n_s ~
        \sqrt{1+ \left({R^2\over r^2 \bar{A}}\partial_{\tilde{\sigma}}r
               \right)^2}\ , \label{W-energy}
\eeq
\beq
  \tilde{\sigma}=a_0(t)\int d\sigma\gamma(\sigma)
       =a_0(t)\int d\sigma {1\over 1-\sigma^2/4}\, ,
\eeq
\beq 
 n_s=\left({r \over R}\right)^2 \bar{A}\bar{n}\, , \label{matter-ns0} 
\eeq
In the present case, we use
the proper coordinate $\tilde{\sigma}$ instead of the comoving coordinate ${\sigma}$ 
to measure the distance between the quark and anti-quark. 

\vspace{.5cm}
It is the criterion of the confinement that $n_s$ has a finite minimum value at some appropriate
$r(=r^*)$. Actually, in such a case, $E$ is approximated as \cite{GIN1} 
\beq
 E\sim {n_s(r^{*})\over 2\pi \alpha'} {L}\ , \label{linear-P}
\eeq 
where 
\beq
  {L}=2\int_{\tilde{\sigma}_{min}}^{\tilde{\sigma}_{max}}d \tilde{\sigma}\, ,
\eeq
and $\tilde{\sigma}_{min}$ ($\tilde{\sigma}_{max}$) is the value at
$r_{min}$ ($r_{min}$) of the string configuration \cite{GIN2}.
The tension of the linear potential between the quark
and anti-quark is therefore given as
\beq \label{tension2}
 \tau_{q\bar{q}}={n_s(r^{*})\over 2\pi \alpha'}\, .
\eeq

\vspace{.5cm}
\noindent{\bf Two groups of the matter}

\vspace{.3cm}
In the case of the matter $\lambda_m$ considered above, $n_s$ is obtained by using 
Eqs.(\ref{sol-10-3}) and (\ref{sol-11-4}) as follows
\beq\label{matter-An}
  \bar{A}=1-{r_m^2 \over r^2}\, , \quad \bar{n}=1+n{r_m^2 \over r^2}\, ,
\eeq
\beq
   n_s=\left({r \over R}\right)^2 
     \left(1+n{r_m^2 \over r^2}\right) \left(1-{r_m^2 \over r^2} \right) \, \label{matter-ns}.
\eeq
where $r_m^2=\lambda_m R^4/4$. When this matter is dominant, we find the nesessary condition for the 
confinement is given by
\beq\label{group-A}
  {\rm (A)}\qquad \lambda_m<0\, , \quad {\rm and} \quad n<0\, .
\eeq
In this case, we obtain
\bea
   r^*&=&\left(-n\left({\lambda_m R^4\over 4}\right)^2\right)^{1/4}\, , \label{minimum-exotic-2} \\
n_s(r^*)&=&{n\lambda_m R^2\over 4}\left(1+{1\over \sqrt{|n|}}\right)^2\, ,\label{minimum-exotic-1}  
\eea
On the other hand, for the case,
\beq\label{group-B}
  {\rm (B)} \quad {\rm Other ~than ~~(A)} \, ,
\eeq
there is no finite minimum of $n_s$. Therefore, the YM theory is in the deconfinement phase in this
case. As a result, the matter is separated to two groups (A) and (B) as given above. The matter of group
(A) ((B)) works to confine (deconfine) quarks.

\vspace{.3cm}
The above form of (\ref{matter-ns}) is compared to the cases of the positive ($dS_4$)
and negative  ($AdS_4$) constant $\lambda(=\Lambda_4/3)$,
in which the corresponding factors are given as follows
\beq
 n_s^{dS_4}=\left({r \over R}\right)^2 
     \left(1-{r_0^2 \over r^2}\right)^2 \, .
\eeq
and 
\beq
 n_s^{AdS_4}=\left({r \over R}\right)^2 
     \left(1+{r_0^2 \over r^2}\right)^2\, .
\eeq
for each case. Here $r_0^2=\lambda_0 R^4/4$, where $\lambda_0>0$.
We know that the area law of the Wilson loop for the quark and anti-quark is
obtained for the case of $n_s^{AdS_4}$, then we find the confinement. This belongs to the group (A) with
$n=-1$.
On the other hand,
we find the deconfinement phase for $n_s^{dS_4}$ since it has zero valued minimum at $r=r_0$. 
In fact, this belongs to the group (B) with $n=-1$.

\vspace{.5cm}
\noindent{\bf A mixed example}

\vspace{.3cm}
It is possible to consider the case where several kinds of matter with different $n$ and the sign of $\lambda_m$
are coexisting. As a simple example, we consider here the combination of 4D 
negative cosmological constant ($n=-1$) (group (A)) and a matter ($n\neq -1$) of group (B), $\lambda_m>0$.
In this case, 
the combined $\lambda$ is given as
\beq
  \lambda=-\lambda_0+\lambda_m\, ,
\eeq 
where $\lambda_0>0$ and $\lambda_m\propto 1/a_0^{n+1}(t)$. Then $n_s$ is written as
\beq\label{matter-ns-2}
 n_s^{(2)}=\left({r \over R}\right)^2 
     \left(1+{\lambda_0+n\lambda_m \over 4\mu^2r^2}\right) \left(1+{\lambda_0-\lambda_m \over 4\mu^2r^2} \right) \, .
\eeq
In this case, the factor $n_s^{(2)}$ has a finite minimum at $r=r^*$, which is given as
\bea\label{minimum}
  n_s^{(2)}(r^*)&=&{r_c^2+nr_m^2\over R^2}\left(1+\sqrt{{r_c^2-r_m^2\over r_c^2+nr_m^2}}\right)^2\, , \\
  r^*&=&\left((r_c^2+nr_m^2)(r_c^2-r_m^2)\right)^{1/4}\, , \\
\eea
where
\beq
  r_c^2={\lambda_0\over 4\mu^4}\, , \quad r_m^2={\lambda_m\over 4\mu^4}\, .
\eeq
Here (\ref{minimum}) is obtained for $r_c^2>r_m^2$. On the other hand, for $r_c^2<r_m^2$,
the minimum of $n_s^{(2)}$ is zero which is obtained at $r^2=r_m^2-r_c^2$. Then the quark
deconfining phase is realized in this case.


\begin{figure}[htbp]
\vspace{.3cm}
\begin{center}
\includegraphics[width=10.0cm,height=7cm]{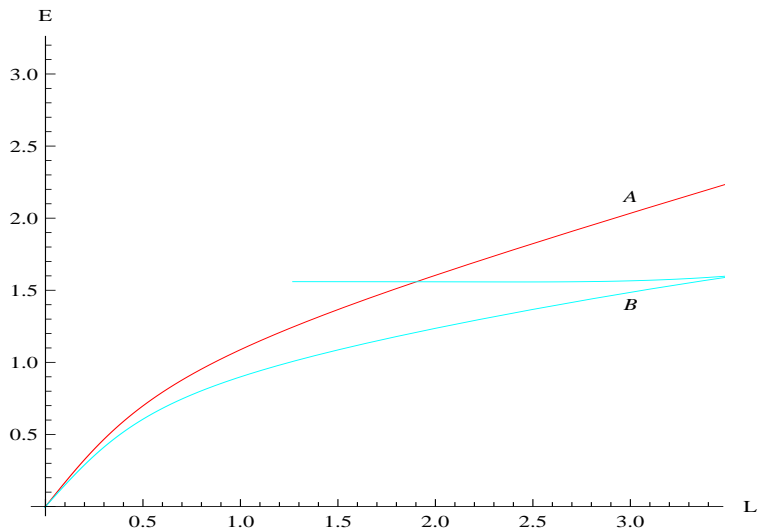}
\caption{Plots of $E$ vs ${L}$ for (A) $\lambda=\lambda_0-0.5$ and (B) $\lambda=\lambda_0+1.5$, where
$n=10$, $\lambda_0=1.0$ and $\mu=1/R=1$.
Further, $\alpha'=1\,$. $E$ increases linearly with ${L}$ at large ${L}$ for the the case of (A), $\lambda-\lambda_0<0$. 
This behavior is
obtained due to negative 4D cosmological constant $-\lambda_0$. The case of (B), $\lambda-\lambda_0>0$, 
shows the case of the
deconfinement phase due to large matter energy density $\lambda$.
\label{wilson-loop}}
\end{center}
\end{figure}

\vspace{.5cm}
In order to understand well the results given above,
the $E-L$ relation and the tension $\tau_{q\bar{q}}$
are examined from numerical estimation in the followings.
Since the Lagrangian in (\ref{ng}) does not explicitly depend on 
the coordinate $\sigma=x$, we find the following relation,
\beq
     {1\over \sqrt{(r/R)^4 \bar{A}^2(r)+(r')^2}}
    \left({r\over R}\right)^4 \bar{n}\bar{A}^2(r)= H\ ,
\eeq
where $H$ denotes a constant of motion. 
And we notice that $r'=\partial_{\tilde{\sigma}}r$.
We can fix $H$ at any point where
we like, so we fix it at $r=r_{min}$. Then, taking as 
$H=\left({r\over R}\right)^2 \bar{n}(r)\bar{A}(r)|_{r_{min}}$, we get
\bea
  &&  {L}=2R^2 \int_{r_{min}}^{r_{\textrm{\scriptsize max}}} dr~
      {1\over r^2 \bar{A}(r)
        \sqrt{r^4 \bar{n}(r)^2\bar{A}(r)^2 /
          \left(r_{min}^4
       \bar{n}(r_{min})^2\bar{A}(r_{min})^2\right)-1}} , \nonumber 
\\
  && E=
   {1\over \pi \alpha'} \int_{r_{min}}^{r_{\textrm{\scriptsize max}}}dr~
   {\bar{n}(r)\over 
     \sqrt{1-r_{min}^4 \bar{n}(r_{min})^2\bar{A}(r_{min})^2/
             \left(r^4 \bar{n}(r)^2\bar{A}(r)^2\right)}} . \label{energy}
\eea
Figure~\ref{wilson-loop} shows the dependence of the energy $E$
on the distance ${L}$ for $\lambda-\lambda_0=-0.5<0$ (curve A) and $\lambda-\lambda_0=1.5>0$  
(curve B). 
In the former case, the matter energy density $\lambda$ is small compared to the 
absolute value of the negative cosmological
constant $\lambda_0$, then we find
the linear potential at large $L$ as expected. On the other hand, in the case of $\lambda >\lambda_0$, we find a 
typical screening behavior
as seen in the finite temperature deconfinement phase. 

From this fact, we could say that the matter of group (B) screens the 
long range force which is needed to confine quarks
and it comes from group (A) matter. Then the effect of the 4D matter of group (B) on the SYM theory is similar
to the one given by the thermal matter in the SYM theory.



\vspace{.5cm}
\subsection{Chiral condensate}

In the next, we study the effects of the decoupled matter on the chiral condensate.
We introduce D7 brane to study the chiral condensate of the quark fields.
The D7-brane action is written by the Dirac-Born-Infeld (DBI) and the
Chern-Simons (CS) terms as follows,
\bea\label{d5action}
S_{D8}&=&-T_{8}\int d^8\xi
 e^{-\Phi}\sqrt{-\det\left(g_{ab}+2\pi\ap F_{ab}\right)}+T_{5}\int \sum_i
\left(\exp^{2\pi\ap F_{(2)}}\wedge c_{(a_1\ldots
a_i)}\right)_{0\ldots 5}~,\\
g_{ab}&\equiv&\p_a X^{\mu}\p_b X^{\nu}G_{\mu\nu}~, \qquad
c_{a_1\ldots
a_i}\,\equiv\,\p_{a_1}X^{\mu_1}\ldots\p_{a_i}X^{\mu_i}C_{\mu_1\ldots\mu_i}~.\nonumber
\eea
where $T_5=1/(\gs(2\pi)^{5}\ls^{6})$ is the brane tension.
The DBI action involves
the induced metric $g_{ab}$ and the $U(1)$ world volume
field strength $F_{(2)}=d A_{(1)}$.

The D7 branes are embedded in the background, which is given by
(\ref{10dmetric-2}) - (\ref{sol-12}), by rewriting
the extra six dimensional of (\ref{10dmetric-2}) as follows
\beq
   \frac{R^2}{r^2} dr^2 +R^2d\Omega_5^2=\frac{R^2}{r^2}\left(
       d\rho^2+\rho^2 d\Omega_3^2+\sum_{i=8}^9{dX^i}^2\right)\, ,
\eeq
where the new coordinate $\rho$ is introduced instead of $r$ with the relation
\beq
  r^2=\rho^2+(X^8)^2+(X^9)^2
\eeq
Thus, the induced metric of the D7 brane is obtained as
\beq\label{8dmetric}
ds^2_{8}={r^2 \over R^2}\left(-\bar{n}^2dt^2+\bar{A}^2a_0^2(t)\gamma^2(x)(dx^i)^2\right)+
\frac{R^2}{r^2} \left(
       (1+{w'}^2)d\rho^2+\rho^2 d\Omega_3^2 \right) \, , 
\label{finite-c-sol-3}
\eeq
where the profile of the D7 brane is taken as $(X^8,X^9)=(w(\rho),0)$
and $w'=\partial_{\rho}w$, then 
\beq
  r^2=\rho^2+{w}^2\, .
\eeq
In the present case,
there is no R-R filed, so the action is given only by the one of DBI as
\beq
  S_{D8}=-T_{8}\Omega_3\int d^4x a_0^3(t)\gamma^3(x)\int d\rho \rho^3
   \bar{A}^3\bar{n}\sqrt{1+{w'}^2(\rho)}\, ,
\eeq 
where $\Omega_3$ denotes the volume of $S^3$ of the D7's world volume.

\vspace{.5cm}
From this action, the equation of motion for $w$ is obtained as
\beq\label{eq-w}
  w''+\left({3\over \rho}+{\rho+ww'\over r}\partial_r (\log (\bar{A}^3\bar{n}))\right)
     w'(1+{w'}^2)-{w\over r}(1+{w'}^2)^2\partial_r ( \log (\bar{A}^3\bar{n}) )=0\, .
\eeq
The constant $w$ is not the solution of this equation, so the supersymmetry is broken.

\vspace{.6cm}
\noindent{\bf For group (A) matter}
\vspace{.3cm}

For the matter of group (A), the numerical solutions of (\ref{eq-w}) for $w(\rho)$ are shown in the Fig. \ref{chiral-c}.
{In general, in this case, we find finite chiral condensate $\langle\bar{\Psi}\Psi\rangle=c$ for any $m_q$ since 
the curves decrease from the above with increasing $\rho$ according to the following asymptotic form}
\beq\label{chiral-br}
  w=m_q+{c\over \rho^2}+\cdots\, ,
\eeq
at large $\rho$ with $c>0$. 
We can observe spontaneous chiral symmetry breaking from the third curve shown in the Fig. \ref{chiral-c}.
It shows the mass generation of a massless quark due to the
chiral condensate $\langle\bar{\Psi}\Psi\rangle$. 

\begin{figure}[htbp]
\vspace{.3cm}
\begin{center}
\includegraphics[width=6.0cm,height=6cm]{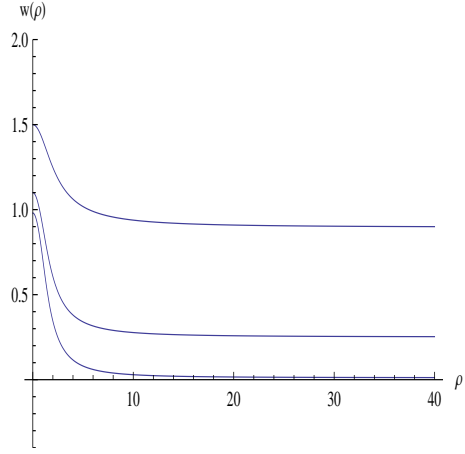}
\caption{An example of group (A) matter. Plots of $w(\rho)$ vs $\rho$ for $\lambda=-2, ~\mu=1$, $n=-0.5$. 
The curves are given for $w(0)=1.5,~1.1,~0.98$ from the top to the bottom. 
\label{chiral-c}}
\end{center}
\end{figure}

As a result, we could say that the spontaneous mass generation of massless quarks
is realized due to the matter of group (A) in the 4D space-time. 
Then the matter of group (A) contributes
to both confinement and chiral symmetry breaking for the SYM theory. The situation is similar to the case of  $AdS_4$.

\vspace{.6cm}
\noindent{\bf For group (B) matter}
\vspace{.3cm}

For the matter of group (B), they are separated further to the two groups by the behavior
near $\rho=0$, which is given as
\beq
  w=w_0+w_2\rho^2+\cdots \, ,
\eeq
for $w_0>r_m$.
\footnote{In the case of $w_0 < r_m$, the value of $\rho$ is restricted as
$\rho\geq\sqrt{r_m^2-w_0^2}$ since the metric $\bar{A}$ has zero point at $r=r_m$. $\bar{n}$ also has zero point
at different value of $r$ , and it is larger than $r_m$ for $n<-1$ and $\lambda_m>0$. However, we do not consider
this case here since it does not produce new things.}
As shown below, we find two cases i) positive and ii) negative $w_2$, the sign of the coefficient $w_2$.
We notice that only the case of negative $w_2$ appears for group (A) matter.

\vspace{.6cm}
{\bf Behavior of $w$ near $\rho=0$}
\vspace{.3cm}

\begin{figure}[htbp]
\vspace{.3cm}
\begin{center}
\includegraphics[width=6.0cm,height=6cm]{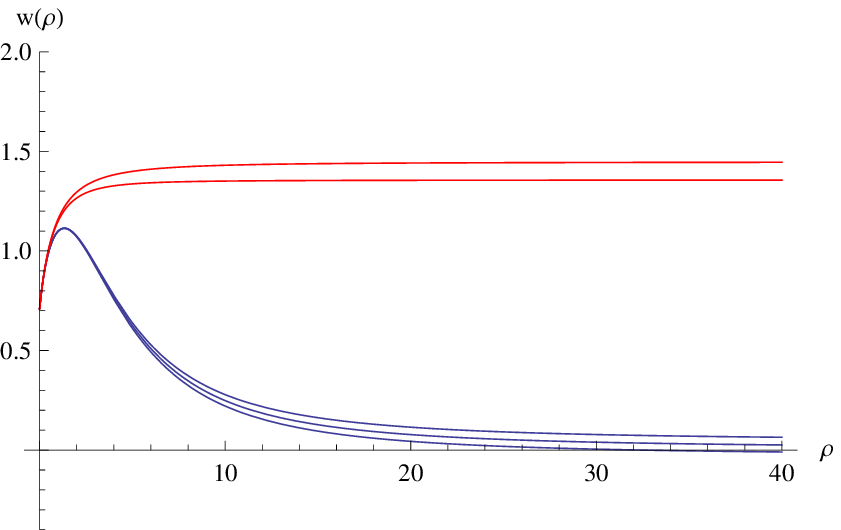}
\includegraphics[width=6.0cm,height=6cm]{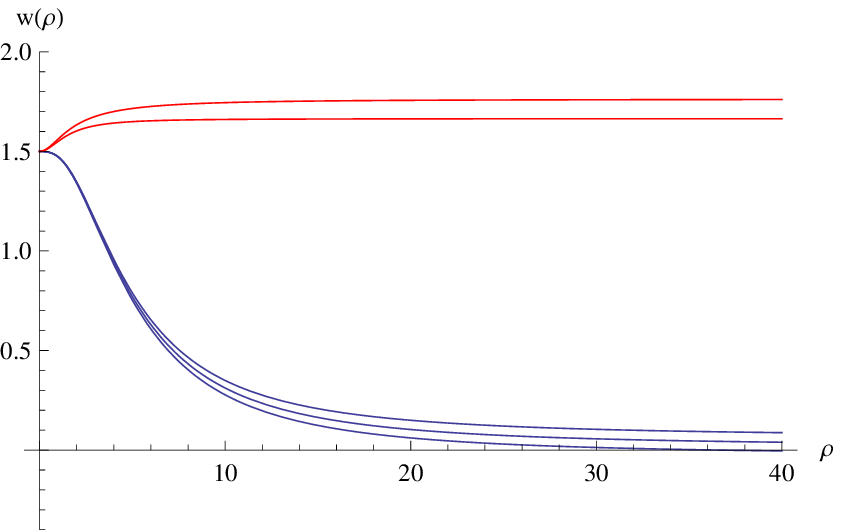}
\caption{For the group (B) matter. Plots of $w(\rho)$ vs $\rho$ for $\lambda=2, ~\mu=1$.
The curves are given for $n=2,~3,~50,~60,~70$ from the top to the bottom. 
The left (right) figure for $r_m<w(0)=0.71<2r_m$
($w(0)=1.50<2r_m$). Each curves of the left and right figures have common behavior near $\rho=0$
\label{chiral-c}}
\end{center}
\end{figure}

The behavior near small $\rho$ is seen as follows.
By substituting this into (\ref{eq-w}) and using (\ref{matter-An}), we find
\beq
  w_2={r_m^2(4nr_m^2+(3-n)w_0^2)\over 4w_0(w_0^2-r_m^2)(w_0^2+nr_m^2)}\, .
\eeq
Then we find $w_2>0$ for the following two cases,
\beq
    w_0<2r_m\, , \quad  n>0\, 
\eeq
and 
\beq
    w_0>2r_m\, , \quad n<{3w_0^2\over w_0^2-4r_m^2}\, .
\eeq
On the other hand, for the following third case,
\beq\label{third-case}
  w_0>2r_m\, , \quad  n>{3w_0^2\over w_0^2-4r_m^2}\, ,
\eeq
we find $w_2<0$. 

In the latter case (\ref{third-case}), we could expect finite chiral condensate
as in the case of the matter of group (A). In order to see this point we need numerical analysis since
it is difficult to obtain the analytic solution which describe $w(\rho)$ for all region of $\rho$.

\vspace{.3cm}
\noindent{\bf (a) For $n=3$ and $n=4$ \\ (non-relativistic matter and radiation)}
\vspace{.3cm}

For $n+1=3$ (for $\rho_m$ or non-relativistic matter) and 4 (for $\rho_r$ or radiation), 
it seems to be impossible to find the spontaneous mass generation even if 
the situation is set as 
$w_2<0$. This is assured from the results shown in the Fig. \ref{chiral-c}, where the two cases
of $n+1=3$ and 4 are shown for both $r_m<w_0<2r_m$ and $2r_m<w_0$ by the upper red curves in each
case. The behavior of these solutions of $w(\rho)$ shows the similar one of the high temperature
or the negative constant $\lambda_0$ case. In both cases, the SYM theory is in the deconfinement
and chiral symmetric phase. In the previous section, we have shown that the role of the group (B) matter
is similar to the one of the high temperature SYM in the deconfinement phase. In this sense,
the results for the matter of $n+1=3$ and 4 is consistent with the results of the previous section.

\vspace{.3cm}
\noindent{\bf (b) Large $n$ matter}

\vspace{.3cm}
Next, we consider the matter given by ${\rho_u/ a_0^{3(1+u)}}={\rho_u/ a_0^{n+1}}$ with large $n$ from the
viewpoint of theoretical interest.
Astonishingly, as shown in the Fig. \ref{chiral-c}, the 
behavior of the solutions for very large $n$ show the asymptotic form given by the Eq.
(\ref{chiral-br}) with positive chiral condensate, $c$. So we summarize the analysis for large $n$ as follows;
\begin{itemize}
\item We could find the solution with positive chiral condensate, $c$,
for any value of $w_0$ for enough large $u$.
\item The solution with $m_q=0$ and $c>0$ could be found at appropriate large value of $n$ for any
matter of group (B). Actually, we could show such solutions at about  $n(\sim 70)$ for both cases of $w_2<0$
and $w_2>0$ as shown in the Fig. \ref{chiral-c}.
\end{itemize}

\vspace{.3cm}
We give a comment on this matter of large $u$ (or $n$). 
Consider a model for this matter in terms of a scalar field $\phi$ with a
self-interacting potential which is given by $V(\phi)$. In this case, the parameter $u$
for this scalar is given as,
\beq
 u={{1\over 2}\dot{\phi}^2-V(\phi)\over {1\over 2}\dot{\phi}^2+V(\phi)}\, ,
\eeq
where $\phi(t)$ is assumed to be solved for an appropriate form of $V(\phi)$. The present purpose is not to
study this model in detail but to point out the possibility of large $u$ case. This may be found when $V$ becomes
negative at some time-interval of varying $\phi(t)$. It would be an interesting problem
how and when this interval would appear in the universe. 
However we will discuss this point in other article where cosmological
problem is studied at the same time.


\vspace{.5cm}
\section{D5 Branes and Baryon}
Here we show the stability of the baryon vertex through the analysis of the D5 brane action
which corresponds to the baryon vertex, which combines
the $N_c$ quarks to make a color singlet. 

Firstly, we briefly review the model based on type IIB superstring theory \cite{wittenbaryon,imamura,cgs,cgst,GI}.
In the type IIB model, the vertex is described
by the D5 brane which wraps $S^5$ of the 10D manifold $M_5\times S^5$. In this case, in the bulk, there exists 
the following form of self-dual Ramond-Ramond field strength
\bea\label{fiveform}
G_{(5)}&\equiv &dC_{(4)}={4\over R}\left(\epsilon_{S^5}+{}^*\epsilon_{S^5}\right)\, \\
            \epsilon_{S^5} &=&R^5\mbox{vol}(S^5) d\theta_1\wedge\ldots\wedge d\theta_5
\eea 
where $\mbox{vol}(S^5)\equiv\sin^4\theta_1
\mbox{vol}(S^4)\equiv\sin^4\theta_1\sin^3\theta_2\sin^2\theta_3\sin\theta_4$,
and  $\epsilon_{S^5}$ denotes the volume form of $S^5$ part. 
The flux from the stacked D3 branes flows into the D5 brane
as $U(1)$ field which is living in the D5 brane. 

\vspace{.5cm}
The effective action of D5 brane is given by using the Born-Infeld and
Chern-Simons term as follows
\begin{eqnarray}\label{d5action}
S_{D5}&=&-T_{5}\int d^6\xi
 e^{-\Phi}\sqrt{-\det\left(g_{ab}+2\pi\ap F_{ab}\right)}+T_{5}\int
\left(2\pi\ap F_{(2)}\wedge c_{(4)}\right)_{0\ldots 5}~,\\
g_{ab}&\equiv&\p_a X^{\mu}\p_b X^{\nu}G_{\mu\nu}~, \qquad
c_{a_1\ldots
a_4}\,\equiv\,\p_{a_1}X^{\mu_1}\ldots\p_{a_4}X^{\mu_4}C_{\mu_1\ldots\mu_4}~.\nonumber
\end{eqnarray}
where $T_5=1/(\gs(2\pi)^{5}\ls^{6})$ and  $F_{(2)}=d A_{(1)}$, which represents the $U(1)$ worldvolume
field strength.
In terms of (the pullback of)
the background five-form field strength $G_{(5)}$, the above action can be rewritten as
$$
S_{D5} = -T_5 \int d^6\xi~ e^{-\Phi}
     \sqrt{-\det(g+F)} +T_5 \int A_{(1)}\wedge G_{(5)}~,
$$

The embedding of the D5 brane is performed by solving the $r(\theta)$, $x(\theta)$, and $A_{(1)}(\theta)$
\cite{GI}.
They are retained as dynamical fields in the D5 brane action as the function of $\theta\equiv\theta_1$ only. 
The equation of motion for the gauge field $A_{(1)}$ is written as
$$
\partial_\theta D = -4 \sin^4\theta,
$$
where the dimensionless displacement 
is defined as the variation of the
action with respect to $E=F_{t\theta}$, namely
$D=\delta \tilde{S}/\delta F_{t\theta}$ and 
$\tilde{S}=S/T_5 \Omega_{4}R^4$. The solution to this equation 
is
\beq \label{d}
D\equiv D(\nu,\theta) = \left[{3\over 2}(\nu\pi-\theta)
  +{3\over 2}\sin\theta\cos\theta+\sin^{3}\theta\cos\theta\right].
\eeq
Here, the integration constant $\nu$ is expressed as $0\leq\nu=k/N_c\leq 1$,
where $k$ denotes the number of Born-Infeld strings emerging from one of the pole of
the ${\bf S}^{5}$. 

\vspace{.5cm}
Next, it is convenient to eliminate the gauge 
field
in favor of $D$, then the Legendre transformation is performed for the original Lagrangian to
obtain an energy
functional as \cite{cgs,cgst,GI}:
\beq \label{u}
U = {N\over 3\pi^2\alpha'}\int d\theta~\bar{n}
\sqrt{r^2+r^{\prime 2} +(r/R)^{4}x^{\prime 2}(\bar{A}a_0\gamma)^2}\,
\sqrt{V_{\nu}(\theta)}~.
\eeq
\beq\label{PotentialV}
V_{\nu}(\theta)=D(\nu,\theta)^2+\sin^8\theta
\eeq
where we used $T_5 \Omega_{4}R^4=N/(3\pi^2\alpha')$. Then, in this expression, (\ref{u}), 
$r(\theta)$ and $x(\theta)$ are remained, and they are solved by minimizing $U$. As a result,
the D5 brane configuration is determined. 

For the simplicity, here, we restrict to the point like configuration, namely $r$ and $x$ are constants.
In this case, we have for the matter considered here
\beq \label{u-p}
U = ~r\bar{n}(r) U_0=r\left(1+n{\lambda_m \over 4\mu^2}\left({R\over r}\right)^2\right)U_0\, , 
\eeq
where $U_0$ is a constant given as
\beq \label{u-0}
  U_0= {N\over 3\pi^2\alpha'}\int d\theta \sqrt{V_{\nu}(\theta)}~.
\eeq
From (\ref{u-p}), we find that $U$ has a minimum at $r_m=\sqrt{n\lambda_m}R^2/2$. For group (A), this assures
the stability of the baryon vertex.

As for the group (B) matter, we must restrict $r$ as $r>\sqrt{\lambda_m}R^2/2$. So we can see the stability
for $n>1$. This situation is different from
the effect of the thermal Yang-Mills field which screens the long range confinement force and destabilize
the baryon vertex \cite{GSUY}. It would be an interesting problem to study the stability for more complicated case,
however, it will be postponed to the future work here.


\vspace{.5cm}
\section{Summary and Discussions}

Here, the holographic {approach is extended to the} SYM theory in 
{a time-dependent curved space-time. Then the SYM theory in such a} universe, 
which is represented by the RW type of metric,
is examined by adding the 4D cosmological constant and the matter which is decoupled from the SYM field
but control the RW metric. 
In the holographic approach, the SYM theory decouples also from the gravity on the boundary. 
Then, the form of the RW metric on the boundary is given as a solution of 4D Friedmann equation, which is
obtained from the Einstein equation on the boundary with the added matter. The matter therefore determines the
scale factor of the RW metric. This process to solve the 4D Friedmann equation
is performed independently of the SYM theory. 
In spite of this fact, the dynamical properties of the SYM
theory are controlled by the added matter through the holographic dual 5D Einstein equation,
{which leads to a deformed AdS$_5$. This background is denoted as $\widetilde{AdS_5}$.}
{Namely, the effect of the 
matter added is reflected to 
the bulk geometry $\widetilde{AdS_5}$, which is deeply related to the 4D Friedmann equation as a result.}

The energy density of various kinds of matter is introduced in the form of  
$\rho_u/a_0^{3(u+1)}(t)$ $\propto 1/a_0^{n+1}(t)$
in the 4D Friedmann equation. For example, the non-relativistic
matter and the radiation are correspond to the one of $u=0$ and $u=1/3$ respectively. 
The effects of these ordinary kinds of matter are similar to the case of the positive
cosmological constant, namely we find the similar dynamical properties to the one found for the SYM
theory in the dS$_4$.
In these cases, therefore, the confinement force is screened above an appropriate distance, 
and furthermore the chiral symmetry is restored.

On the other hand, we find confinement and chiral symmetry breaking for $\rho_u<0$ and $n<0$
(group(A)), and the negative cosmological
constant belongs to this group. Then, we can separate the matter
to two groups (A) and others (group (B)). 
As for the group (B), they are further separated to ordinary kinds of matter and exotic one. 
The ordinary kinds of matter are the one of $u=0$ and $u=1/3$ mentioned above, and the positive
cosmological constant. They lead to deconfinement and chiral symmetric phase of the SYM theory.
The one called as exotic matter has very large $u$ (or large $n$) compared to the ordinary kinds of matter.
As an effect of this matter, the chiral symmetry of the theory is spontaneously broken. 

Furthermore we have examined the effect of these kinds of matter
on the D5 brane as the baryon vertex, which is in general
unstable in the case of deconfining phase. However, in the case of the matter of group (B)
with any $n\geq 1$, we find that the
D5 brane is stable. 

The results obtained here are important when we study the developing
universe since the matter belonging to the SYM theory is the important ingredient of the 
universe, and the SYM is surrounded by other {kind of matter which control the dynamical properties of
the SYM theory.}

\vspace{.3cm}
\section*{Acknowledgments}

One of the authors, K. G. , thanks R. Meyer for useful discussions at early stage of this work.

\vspace{1cm}

\newpage
\end{document}

%% file: macros.tex
\newcommand{\beqs}{\begin{equation*}}

\newcommand{\eeqs}{\end{equation*}}

\newcommand{\beqas}{\begin{eqnarray*}}
\newcommand{\beqa}{\begin{eqnarray}}

\newcommand{\eeqas}{\end{eqnarray*}}
\newcommand{\eeqa}{\end{eqnarray}}

\newcommand{\blist}{\begin{itemize}}

\newcommand{\elist}{\end{itemize}}

\providecommand{\href}[2]{#2}

\DeclareFontFamily{OT1}{rsfs}{}
\DeclareFontShape{OT1}{rsfs}{m}{n}{ <-7> rsfs5 <7-10> rsfs7 <10->rsfs10}{} 
\DeclareMathAlphabet{\mycal}{OT1}{rsfs}{m}{n}

\def\TrL2{{\rm Tr}_{L^2}}

\def\atbdry{\Big|_{\partial \cM}}
\def\atbdry0{\Big|_{\partial \cM_0}}
\def\atbdry1{\Big|_{\partial \cM_1}}








%% file: holo-cos-14-G.bbl
\begin{thebibliography}{99}

\bibitem{MGW}
J.~M.~Maldacena,
Adv.\ Theor.\ Math.\ Phys.\  {\bf 2}, 231 (1998) [hep-th/9711200].

S.~S.~Gubser, I.~R.~Klebanov and A.~M.~Polyakov,
Phys.\ Lett.\ B {\bf 428}, 105 (1998) [hep-th/9802109].

E.~Witten,
Adv.\ Theor.\ Math.\ Phys.\  {\bf 2}, 253 (1998) [hep-th/9802150].
A.M. Polyakov, Int. J. Mod. Phys. {\bf A14} (1999) 645,
        ({\tt hep-th/9809057}).
        
\bibitem{KK}
A.~Karch and E.~Katz, JHEP {\bf 0206}, 043(2003)  [hep-th/0205236].
\bibitem{KMMW}
M.~Kruczenski, D.~Mateos, R.C.~Myers and D.J.~Winters, JHEP {\bf 0307}, 049(2003)  [hep-th/0304032].
\bibitem{KMMW2}
M.~Kruczenski, D.~Mateos, R.C.~Myers and D.J.~Winters, [hep-th/0311270].
\bibitem{Bab}
J.~Babington, J.~Erdmenger, N.~Evans, Z.~Guralnik and I.~Kirsch, 
hep-th/0306018. 
\bibitem{ES}
N.~Evans, and J.P.~Shock, 
hep-th/0403279. 
\bibitem{SS} T.~Sakai and J.~Sonnenshein, [hep-th/0305049].
\bibitem{NPR}
C.~Nunez, A.~Paredes and A.V.~Ramallo, JHEP {\bf 0312}, 024(2003)  
[hep-th/0311201].
\bibitem{GY} K. Ghoroku and M. Yahiro,  Phys.\ Lett.\ B {\bf 604}, 235(2004),
[hep-th/0408040].
\bibitem{CNP} R. Casero, C. Nunez and A. Paredes,
 Phys.Rev. D73 (2006) 086005.
 

\bibitem{Hawking}
  S.~Hawking, J.~M.~Maldacena and A.~Strominger,
  JHEP {\bf 0105} (2001) 001
  [arXiv:hep-th/0002145].

\bibitem{Alishahiha1}
  M.~Alishahiha, A.~Karch, E.~Silverstein and D.~Tong,
  AIP Conf.\ Proc.\  {\bf 743} (2005) 393
  [arXiv:hep-th/0407125].
\bibitem{Alishahiha2}
  M.~Alishahiha, A.~Karch and E.~Silverstein,
  JHEP {\bf 0506} (2005) 028
  [arXiv:hep-th/0504056].

\bibitem{H}
T. Hirayama, JHEP {\bf 0606}, 013(2006)  [hep-th/0602258].
\bibitem{GIN1}
K.~Ghoroku M.~Ishihara and A.~Nakamura, Phys. Rev. {\bf D74} 124020 (2006) .
\bibitem{AIS} S.J. Avis, C.J. Isham and D. Storey,
    Phys. Rev. {\bf D 18}, 3565 (1978).

\bibitem{GIN2}
K.~Ghoroku M.~Ishihara and A.~Nakamura, Phys. Rev. {\bf D75} 046005 (2007) .
\bibitem{GSUY} K. Ghoroku, T. Sakaguchi, N. Uekusa and M. Yahiro,
 Phys. Rev. {\bf D71}(2005)106002.
\bibitem{Gub} S. Gubser, [hep-th/9902155]
 
\bibitem{EGR} J. Erdmenger, K. Ghoroku, R. Meyer, 
"Holographic (De)confinement Transitions in Cosmological Backgrounds ", 
Phys.Rev.D84:026004,2011,
[arXiv:1105.1776 (hep-th)] 

\bibitem{EGR2} J. Erdmenger, K. Ghoroku, R. Meyer, Ioannis Papadimitriou,
 "Holographic Cosmological Backgrounds, Wilson Loop (De)confinement and Dilaton Singularities"
[arXiv:1205.0677 (hep-th)] 

\bibitem{BDEL} P. Binetruy, C. Deffayet, U. Ellwanger and D. Langlois,
Phys.Lett. B477 (2000) 285-291,[hep-th/9910219]
\bibitem{Lang} D. Langlois, hep-th/0005025, 0306281

\bibitem{KS2}
A.~Kehagias and K.~Sfetsos, Phys.\ Lett.\ B {\bf 456}, 22(1999) 
[hep-th/9903109]. 
\bibitem{LT}
H.~Liu and A.A.~Tseytlin [hep-th/9903091].
\bibitem{GGP} G. W. Gibbons, M. B. Green and M. J. Perry,
  Phys.Lett. B370 (1996) 37-44, [hep-th/9511080].
  


\bibitem{KSS} 
S.~de Haro, S.~N.~Solodukhin and K.~Skenderis,
Commun.\ Math.\ Phys.\  {\bf 217}, 595 (2001)
[hep-th/0002230].
\bibitem{BFS} Massimo Bianchi, Daniel Z. Freedman, Kostas Skenderis,
Nucl.Phys. B631 (2002) 159-194, [hep-th/0112119].
\bibitem{FG}
{C. Fefferman and C. Robin Graham, `Conformal Invariants', in 
{\it Elie Cartan et les Math\'ematiques d'aujourd'hui} 
(Ast\'erisque, 1985) 95.}

\bibitem{NS} Shin Nakamura and Sang-Jin Sin, JHEP0609:020,2006, [hep-th/0607123].

\bibitem{Duff93} M.J. Duff, Class.Quant.Grav. 11 (1994) 1387-1404,[hep-th/9308075]
\bibitem{birrell} N.D. Birrell and P.C.W. Davies, "Quantum fields in curved space"
1982, Cambridge Univ. Press.


\bibitem{wittenbaryon}
E.~Witten, ``Baryons and Branes in Anti de Sitter Space,''
J.~High Energy Phys. {\bf 07} (1998) 006,
\href{http://xxx.lanl.gov/abs/hep-th/9805112}{{\tt hep-th/9805112}}.

\bibitem{imamura}
Y.~Imamura, ``Supersymmetries and BPS Configurations on Anti-de 
Sitter Space,''
Nucl.~Phys. {\bf B537} (1999) 184,
\href{http://xxx.lanl.gov/abs/hep-th/9807179}{{\tt hep-th/9807179}}.

\bibitem{cgs}
C.~G.~Callan, A.~G\"uijosa, and K.~Savvidy,
``Baryons and String Creation from the Fivebrane Worldvolume 
Action,''
\href{http://xxx.lanl.gov/abs/hep-th/9810092}
{{\tt hep-th/9810092}}.

\bibitem{cgst}
  C.~G.~Callan, A.~G\"uijosa, K.~G.~Savvidy and O.~Tafjord,
  ``Baryons and flux tubes in confining gauge theories from brane actions,''
  Nucl.\ Phys.\ B {\bf 555} (1999) 183
  [arXiv:hep-th/9902197].

\bibitem{GI}
  K.~Ghoroku, M.~Ishihara,
  ``Baryons with D5 Brane Vertex and $k$-quarks states,''
  Phys.\ Rev.\  D {\bf 77}, 086003 (2008)
  {\tt [arXiv:hep-th/0801.4216]}.\\
  K.~Ghoroku, M.~Ishihara, A. Nakamura and F. Toyoda,
  ``Multi-quark baryon and color screening at finite temperature,''
  Phys.\ Rev.\  D {\bf 79}, 066009 (2009)
  {\tt [arXiv:hep-th/0806.0195]}.




\end{thebibliography}
